\documentclass[twocolumn,aps,prl,showpacs,floatfix]{revtex4}

\usepackage{amsmath}
\usepackage{epsfig}
\usepackage{bm}
\usepackage{color}
\newcommand{\B}[1]{{\bm{#1}}}

\newcommand{\C}[1]{{\mathcal{#1}}}
\usepackage[latin1]{inputenc}
\newcommand{\be}{\begin{equation}}
\newcommand{\ee}{\end{equation}}
\newcommand{\bea}{\begin{eqnarray}}
\newcommand{\eea}{\end{eqnarray}}

\usepackage{amsmath}
\usepackage{amsfonts}
\usepackage{amssymb}
\usepackage{bm}
\usepackage{color}
\usepackage{graphicx}

\newcommand{\la}{\langle}
\newcommand{\ra}{\rangle}

\def\Re{\ensuremath{{\C R}\mkern-3.1mu e}}
\def\De{\ensuremath{{\C D}\mkern-3.1mu e}}

\begin{document}

\title{Comparison of Theory and Direct Numerical Simulations of Drag Reduction by Rodlike Polymers in Turbulent Channel Flows}
\author{Roberto Benzi$^1$, Emily S. C. Ching$^2$, Elisabetta De Angelis$^3$ and Itamar Procaccia$^4$}
\affiliation{$^1$Dipartimento di Fisica and INFN,
  Universit\`a di Roma ``Tor Vergata'', Via della Ricerca Scientifica
  1, 00133 Roma, Italy\\
 $^2$Dept. of Physics and Inst. of Theoretical Physics, 
The Chinese University of Hong Kong, Shatin, Hong Kong\\
 $^3$ Dip. di Meccanica e Areonatuitca, Via Eudossiana 18, I-00184 Rome, Italy\\
 $^4$ Department of Chemical Physics, The Weizmann Institute
of Science, Rehovot 76100, Israel}

\date{\today}
 
\begin{abstract}
Numerical simulations of turbulent channel flows, 
with or without additives, are limited in the extent
of the Reynolds number \Re~and Deborah number \De. The comparison of such simulations to theories of drag reduction, which are usually derived for asymptotically high \Re~ and \De,  calls for some care. In this paper we present a study of drag reduction by rodlike polymers in
a turbulent channel flow using direct numerical simulation and illustrate how these
numerical results should be related to the recently developed theory.
\end{abstract}
\pacs{xxxx}
\maketitle

\section{Introduction}
Drag reduction in wall-bounded turbulent flows can be achieved with the addition of 
either flexible or rodlike polymers \cite{Virk,Bonn}. 
The theory of drag reduction by either type of polymers is by now well established \cite{RMP}. 
The theory for flexible polymers is supported by both experiments and numerical
simulations while that for rodlike polymers has been compared mainly 
with laboratory experiments. More scant are numerical simulations of drag reduction 
by rodlike polymers, and some results became available only quite recently \cite{fibres}. 
Comparisons of theory and simulations in this case must be done with care, since the
available theory assumes high enough Reynolds \Re~and Deborah numbers \De, in contrast to the
situation in simulations where these crucial characteristic numbers are relatively low. 
To clarify the relation between the theory and the available numerical simulations, 
we present in this note a comparison between numerical simulations and theoretical 
predictions adapted to the limited values of \Re~and \De~. 

The equations of motion used for the numerical simulations are
\begin{equation}
\frac{\partial U_i}{\partial t} + U_j \frac{\partial}{\partial x_j} U_i =
- \frac{\partial p}{\partial x_i} + \nu_0 \frac{\partial^2 U_i}{\partial x_j \partial x_j} 
+ \frac{\partial \sigma_{ij}}{\partial x_j}  \ , 
\label{modifiedNS}
\end{equation}
supplemented by the incompressibility constraint $\partial U_i/\partial x_i=0$
where $\B U$ is the velocity field, $p$ is the pressure, the units are chosen such that 
the fluid density is unity, $\nu_0$ is the kinematic viscosity of the neat fluid, 
and $\sigma_{ij}$ is the additional stress tensor due to the rodlike polymers.

The rodlike polymers are represented by rigid and neutrally buoyant elongated particles. 
The particles are assumed to be massless and have no inertia.
The orientation of each polymer is given by a unit vector $\B n$. 
In turbulent flows with strong shear such that thermal Brownian rotations can be neglected, 
the evolution equation for the second
moment of the conformation tensor ${\cal R}_{ij} = \langle n_i n_j \rangle$ is given
by~\cite{DoiEdwards}:
\be
\frac{\partial {\cal R}_{ij}}{\partial t} + U_k \frac{\partial {\cal R}_{ij}}{\partial x_k} 
= {\cal S}_{ik} {\cal R}_{kj} + {\cal S}_{jk} {\cal R}_{ki} - 2 {\cal S}_{kl} {\cal R}_{ijkl}  
\label{evolution}
\ee
where ${\cal R}_{ijkl} = \langle n_i n_j n_k n_l \rangle $
and ${\cal S}_{ij}$ are the velocity gradient ${\cal S}_{ij} = {\partial u_i}/{\partial x_j} $.
Moreover, in this non-Brownian limit, the tensor $\sigma_{ij}$ in Eq. (\ref{modifiedNS}) is given by~\cite{DoiEdwards,additive},
\be
\sigma_{ij} = 6 \nu_p S_{kl} {\cal R}_{ijkl} \ ,
\label{sigma}
\ee
where $\nu_p$ is the polymeric contribution to the viscosity at
vanishingly small and time-independent shear, and is proportional to the
product of $\nu_0 \phi$ where $\phi$ is the volume fraction of the
polymers. In \cite{DoiEdwards} it was argued that these equations can be closed
by the simple closure ${\cal R}_{ijkl} = {\cal R}_{ij} {\cal R}_{kl}$.
%%%%%%%%%%%%%%
\section{Simulations}
The equations of motion are numerically
integrated for a channel flow . The dimensions of the integration domain are $2\pi L \times
2L \times 1.2 \pi L$ in the $x$ (streamwise), $y$ (wall-normal) and 
$z$ (spanwise) directions respectively, with $L$ being the channel half-width. 
The numerical formulation is a standard pseudospectral method with
Fourier expansion in directions parallel to the wall and Chebyshev in the direction normal 
to the wall. The grid used is $128 \times 193 \times 64$. 
The direct numerical simulations (DNS) were performed 
at a nominal $\Re \equiv U_0 L/\nu_0$ of 10000 for both the 
Newtonian flow (with  $\eta_p \equiv 6 \nu_p/\nu_0 = 0$) and for
turbulent flow with rodlike polymers with $\eta_p = 25$, where $U_0$ is the mean velocity
at the center of the channel. 
Both flows were forced on average with the same pressure drop 
$p'\equiv -\partial p/\partial x$, so the resulting Reynolds
number based on the friction velocity is the same and equal to 
$\Re_{\tau} \equiv \sqrt{p'L}L/\nu_0= 300$.
In channel geometry the only non-vanishing mean velocity component is 
$V(y) \equiv \langle U_x\rangle$. Accordingly we separate the velocity field into 
its mean and fluctuation, $\B U= V\hat x+\B u$. Below we use the wall units:
$y^+ \equiv y  \Re_\tau/L$ and  $V^+(y^+) \equiv V/\sqrt {p'L}$.

\begin{figure}
\centering
\epsfig{width=.48\textwidth,file=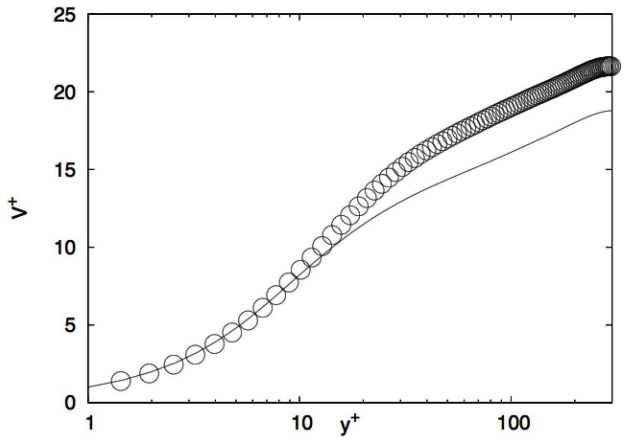}
\caption{Mean streamwise velocity profile $V^+(y^+)$
for both the Newtonian flow (solid line) and the rodlike polymer laden flow (circles).}
\label{fig1}
\end{figure}

In the polymer laden flow, the mean velocity profile as a function of the distance from
the wall (in wall units) exhibits an increase with respect to the Newtonian 
flow~(see~Fig.~\ref{fig1}). This is the phenomenon of drag reduction. Note that the 
relative smallness of \Re~ means here that the
effect is not large, and by $y^+=80$ the velocity profile of the polymer laden flow 
is already parallel to that of the Newtonian flow. So any comparison with 
the theory of drag reduction should be limited to the rather narrow
window of $20<y^+<80$. One should note the very different situations here and 
when \Re~and \De~are very large. In the latter case, 
the mean velocity profile attains the maximum drag reduction asymptote (MDR) and 
never becomes parallel again to the von K\`arm\`an log-law of the Newtonian flows. 

%%%%% FIGURE 1 %%%%%%%%%%%%%%%%%%
%%%%%%%%%%%%%%%%%%%%%%%%%%%%%%%%%%

\begin{figure}
\centering
\epsfig{width=.48\textwidth,file=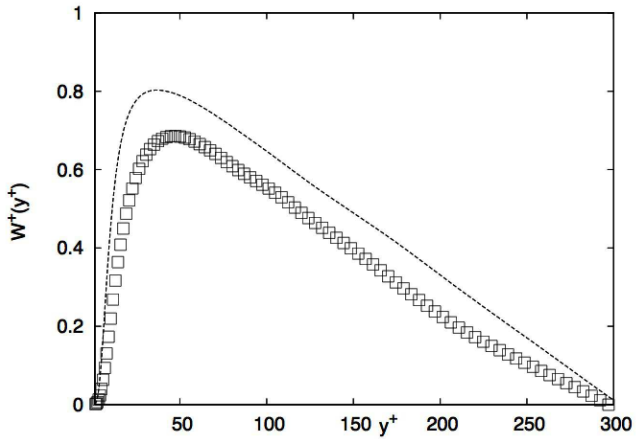}
\caption{The Reynolds stress $W(y)$ for 
the Newtonian flow (dashed line) and the flow with rodlike polymers (squares).}
\label{fig2}
\end{figure}

In Fig.~\ref{fig2}, we show the momentum fluxes. There is a significant reduction in the
Reynolds stress $W(y)\equiv -\langle u_x u_y\rangle$ for the rodlike polymer laden flow 
as compared to the Newtonian flow. The Reynolds stress is the mean mechanical momentum 
flux from the fluid to the wall. It had been explained before that the reduction in the 
momentum flux is at the heart of the mechanism for drag reduction \cite{04LPPT}. Let us 
note that Figs.~\ref{fig1} and \ref{fig2} are quite close to what one observes in 
turbulent channel flows with flexible polymers: an increase of the mean velocity profile 
and a marked decrease of the mean momentum fluxes.

%%%%% FIGURE 2  %%%%%%%%%%%%%%%%%%
%%%%%%%%%%%%%%%%%%%%%%%%%%%%%%%%%%

\section{Comparison of theory with simulations}
\begin{figure}
\centering
\epsfig{width=.5\textwidth,file=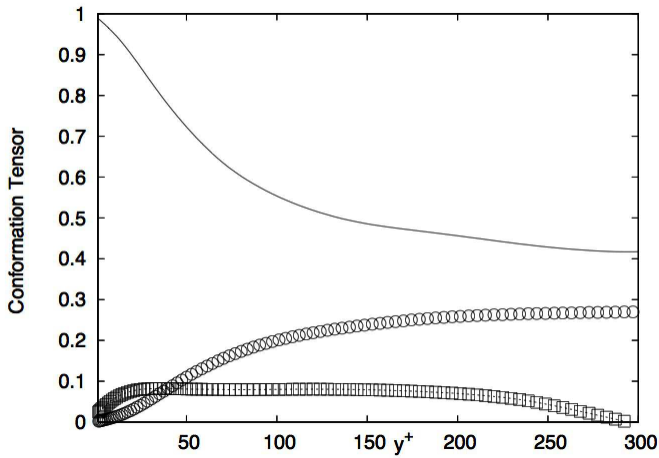}
\caption{The profiles of the averages of the components of the conformation 
tensor $R_{xx}$~(solid line), $R_{yy}$~(circles), $R_{zz}$~(triangles),
and $R_{xy}$~(squares).}
\label{Rcomp}
\end{figure}

\begin{figure}
\centering
\epsfig{width=.50\textwidth,file=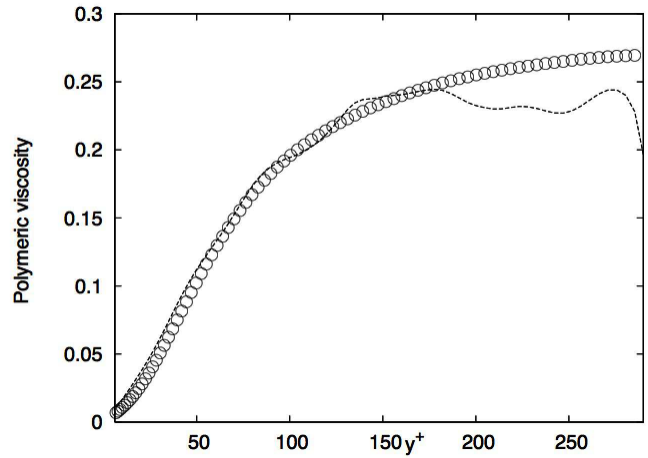}
\caption{A comparison of $R_{yy}(y)$ (dashed line) with 
$\langle \sigma_{xy} \rangle / c_1 \nu_p S(y)$~(circles).}
\label{fig4}
\end{figure}

The theory of turbulent drag reduction by rodlike
polymers~\cite{additive} is based in part on the exact momentum balance equation:
\be
\langle \sigma_{xy} \rangle + \nu_0 S + W = p' (L-y)
\label{momentum}
\ee
where $S(y) = dV(y)/dy$ is the mean shear. A central
ingredient of the theory is the statement that
the polymer contribution to this equation, i.e. $\langle \sigma_{xy}\rangle$, 
can be evaluated as:
\be
\langle \sigma_{xy} \rangle \approx c_1 \nu_p R_{yy}(y) S(y)
\label{eq44}
\ee
with some constant $c_1$ and $R_{ij}=\langle {\cal R}_{ij}\rangle$. 
Similarly, in the energy balance equation:
\be
\nu_0 \langle s_{ij} s_{ij} \rangle + \langle \sigma_{ij} s_{ij} \rangle \approx S W
\label{energy}
\ee
the polymer contribution to the dissipation, denoted here as $\epsilon^p = \langle
\sigma_{ij} s_{ij} \rangle$,  can be evaluated as
\be
\epsilon^p(y) \approx c_2 \nu_p R_{yy}(y) \frac{K(y)}{y^2}
\label{eq58}
\ee
with some constant $c_2$, where $K(y)\equiv \langle |\B u|^2\rangle/2$ 
is the kinetic energy of the fluctuating velocity.
Here $s_{ij}$ is the fluctuating part of the velocity gradient tensor ${\cal S}_{ab}$,
defined by:
\be
{\cal S}_{ab}({\bf r},t) = S(y) \delta_{ax} \delta_{by} + s_{ab}({\bf r},t),
\ \
\langle s_{ab}({\bf r},t) \rangle = 0  .
\ee
It was shown that for large \Re~and \De~these equations predict the establishment of a new
velocity profile, again in the form a power law, but with a considerably larger slope compared
to the Newtonian slope. This asymptotic log-law is known as the Maximum Drag Reduction Asymptote (MDR). Moreover, the existence of a new log-law, with $V^+$ linear in $\log y^+$, 
for the (drag reduced) mean energy profile is directly related
to $R_{yy}(y)$ increasing linearly with $y$.
Physically, the theory states that the effects of the polymer can be treated 
as an $y$-dependent effective viscosity which increases linearly with $y$. 
It is thus of immediate interest to test these predictions also in the present case of
relatively low \Re~and \De. 
To this end, we show in Fig.~\ref{Rcomp} the averages of various components of the 
conformation tensors: $R_{xx}$, $R_{yy}$,  and $R_{xy}$ obtained in the
simulation. We see clearly that $R_{yy}$ increases linearly with $y$ up to $y^+ \sim 80$, 
which is the relevant range where drag reduction takes place in this simulation.
In Fig.~\ref{fig4}, we present the direct comparison of  $\langle \sigma_{xy}\rangle$ 
with $\nu_p R_{yy}(y) S(y)$. The good agreement between the object and its evaluation 
is shown to exceed the region of linearity in $y^+$. Thus two central predictions of 
the theory are well supported by the direct numerical simulation even at the modest 
value of \Re~that is available here. 
%%%%% FIGURE 3 %%%%%%%%%%%%%%%%%%
%%%%%%%%%%%%%%%%%%%%%%%%%%%%%%%%%%

%%%%% FIGURE 4 %%%%%%%%%%%%%%%%%%
%%%%%%%%%%%%%%%%%%%%%%%%%%%%%%%%%%

Needless to say, not every prediction of a theory that is developed as 
an {\em asymptotic} theory in the 
limit of $\Re \to \infty$ and Deborah number $\De \to \infty$ \cite{additive} can be expected
to hold verbatim, and some modification might be required.  In  
the asymptotic theory one argues that 
$R_{xx}\approx 1\gg R_{xy}$, $R_{xy} \gg R_{yy}$ and $R_{yy} \sim R_{xy}^2$. 
We see from Fig, \ref{Rcomp} that this is not the case here. 
To understand these results, note first that 
the Deborah number for flows with rodlike polymers is defined in the literature as 
$\De \equiv S/\gamma_B$,
where $\gamma_B$ is the Brownian rotational frequency.
The simulations were done using $\gamma_B=0$, formally at infinite $\De$ 
for a laminar shear flow. In the present case of a
turbulent channel flow, the effect of turbulence is to induce  
rotations of the polymers, giving rise to an 
effective relaxation frequency $\gamma_{turb}$ which depends 
on the turbulent intensity. Thus the corresponding 
effective Deborah number $\De = S/\gamma_{turb}$ is also finite.
At finite $\Re$ and finite $\De$, the relative sizes of the averages of the 
various components of the conformation tensors need to be reevaluated. 
In the following, we shall show how the theory can be employed
for the case of finite $\Re$ and $\De$ to explain the observed numerical results 
Fig.~\ref{Rcomp}.

We shall start from the equations of motions of the conformation tensor ${\cal R}_{ab}$.
Averaging Eq.~(\ref{evolution}) over the turbulent fluctuations, 
\be
\label{Rab}
\langle U_k \frac{\partial {\cal R}_{ab}}{\partial x_k} \rangle =
SR_{yb}\delta_{ax}+SR_{ya}\delta_{bx} + 2 \delta_{ab} \frac{\Sigma}{3}
- 2R_{ab}(R_{xy}S+\Sigma)
\ee
where $\Sigma=\langle {\cal R}_{ab} s_{ab}\rangle$. 
To derive this equation we first employed the closure assumption
\be
\label{approx}
\langle  {\cal R}_{abcd} {\cal S}_{cd} \rangle \approx
\langle  {\cal R}_{ab} {\cal R}_{cd} {\cal S}_{cd} \rangle
\approx R_{ab}(R_{xy} S +\Sigma) \ .
\ee
The second simplification is the assumption that after removing the mean shear, 
the remaining velocity fluctuations are not too far from isotropic, 
and in the log-layer can also be taken as homogeneous. 
This implies that correlation functions of $s_{ab}$ with ${\cal R}_{ab}$ are
isotropic in space:
\bea
\langle  s_{xc} {\cal R}_{cx} \rangle \approx
\langle  s_{yc} {\cal R}_{cy} \rangle\approx
\langle  s_{zc} {\cal R}_{cz} \rangle &\approx& \frac{\Sigma}{3}\\
\langle s_{ac} {\cal R}_{cb} \rangle &\approx& \delta_{ab} \frac{\Sigma}{3}
\eea
Within the same assumptions we can also offer an approximate evaluation of 
$\Sigma\approx A\sqrt{K/y^2}$ since all the velocity fluctuations are close to 
isotropic. Here $A$ is a constant of the order of unity. 
Finally, we evaluate  $\langle U_k \partial {\cal R}_{ab}/\partial x_k \rangle \approx 0$. This
is seen by integrating by parts and using the the incompressibility constraint on. The
derivative of the average is negligible for fluctuations that are not too far 
from homogeneous. With all these we obtain:
\begin{eqnarray}
\label{1}
&&0= 2SR_{xy} -2SR_{xx}R_{xy}-2\Sigma(R_{xx}-\frac{1}{3}) \\
\label{2}
&&0= SR_{yy}-2R^2_{xy}S-2\Sigma R_{xy} \\
\label{3}
&&0= -2SR_{yy}R_{xy}-2\Sigma(R_{yy}-\frac{1}{3})
\end{eqnarray}
These equations are identical to those obeyed by the conformation tensor 
in a steady laminar shear flow and this tells us that $S/\Sigma$ can be 
taken as the effective {\it y-dependent}
Deborah number in the simulation. Using Eqs.~(\ref{1})-(\ref{3}), we proceed 
to compute the $y$ profiles of $R_{ab}$. To do so, we solve
$R_{ab}$ in terms of $S/\Sigma$. Then we consider the polymers to be small perturbations
and use the momentum and energy balance equations for turbulent Newtonian 
channel flow to get  $S(y)$ and $K(y)$.
Written in wall units, the momentum and energy balance equations for turbulent
Newtonian channel flow read
\bea
S^++W^+ = 1-\frac{y^+}{Re_{\tau}} \\
\delta^2\frac{K^+}{{y^+}^2}+ \frac{{K^+}^{3/2}}{\kappa_k y^+} = W^+S ^+\ .
\eea
where $\kappa_k$ is the
Von Karman constant and $\delta$ is the thickness of the viscous layer, and $\delta
\approx 6$ was found in~\cite{05BDLP}. 

%%%%% FIGURE 5 %%%%%%%%%%%%%%%%%%
\begin{figure}
\hskip  - .1 cm
\epsfig{width=.50\textwidth,file=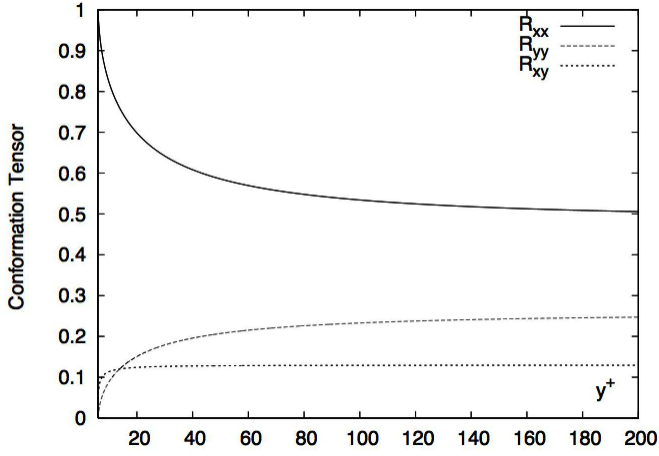}
\caption{ Theoretical prediction of the averages of the 
components of the conformation tensor as a function of the distance from the wall. }
\label{fig5}
\end{figure}
%%%%%%%%%%%%%%%%%%%%%%%%%%%%%%%%%%

In Fig.~\ref{fig5}, we show $R_{xx}$, $R_{xy}$ and $R_{yy}$
obtained by solving (\ref{1}), (\ref{2}) and (\ref{3}) with $A=0.7$. 
We find general agreement
with the results shown in Fig.~\ref{Rcomp}, explaining why in this case the relative
sizes of the averages of the components of the conformation tensor differ from the 
predictions of the asymptotic theory. Our simple modeling can also explain the results 
shown in Fig.~\ref{fig4}.  We obtain from Eq.~(\ref{2}) 
\be
SR_{yy} = 2 R_{xy} (S R_{xy}+\Sigma)
\ee
Using this result and Eqs.~(\ref{sigma}) and (\ref{approx}), we get
\be
\langle \sigma_{xy} \rangle \approx 6 \nu_p R_{xy}(R_{xy}S + \Sigma) = 6 \nu_p R_{yy} S
\ee
Thus, even if the asymptotic theory cannot be applied directly in the present DNS, 
the basic prediction $\sigma_{xy} \approx c_1 \nu_p R_{yy} S$ still holds, 
consistent with the above mentioned approximations, i.e.
neglecting anisotropic contributions like $\la R_{xa}s_{ay}\ra$.

%%%%% FIGURE 8 %%%%%%%%%%%%%%%%%%
%\begin{figure}
%\centering
%\epsfig{width=.40\textwidth,file=sigma_xy.eps}
%\caption{ }
%\label{fig8}
%\end{figure}
%%%%%%%%%%%%%%%%%%%%%%%%%%%%%%%%%%

%A direct check of our conclusion can be seen in Fig.~\ref{fig8} where
%we plot $R_{yy} S$ obtained using by our approximations. A direct comparison
%against Fig.~\ref{fig4} shows that we keep the shape and right of order
%of magnitude of $\sigma_{xy}$.

\begin{figure}
\centering
\epsfig{width=.50\textwidth,file=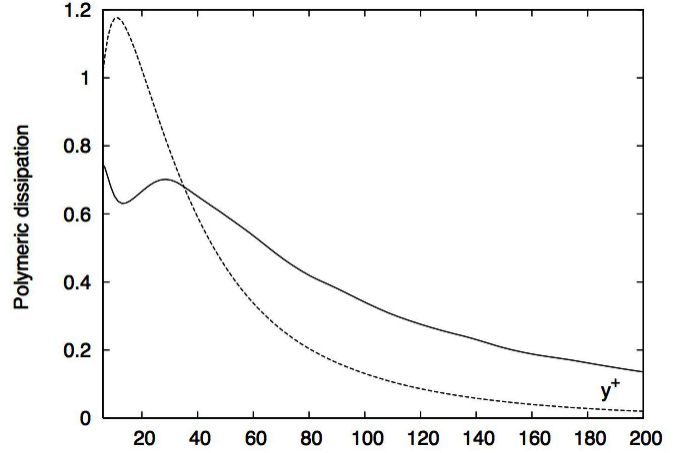}
\caption{Measured~(solid line) and predicted~(dotted line) dissipation 
contributed by the rodlike polymers to the energy balance equation. 
The measured quantity $\epsilon^p$ has two peaks, whereas  $R_{yy}K/y^2$ has only one peak.}
\label{fig9}
\end{figure}

In order to state that $R_{yy}$ is an effective viscosity it should also play the role of
additional viscosity in the energy balance equation. To test the validity of this we
compare the energy dissipation due to polymers $\epsilon^p$ against
the theoretical prediction Eq. (\ref{eq58}). This is done in Fig.~\ref{fig9}.
%%%%% FIGURE 9 %%%%%%%%%%%%%%%%%%
%%%%%%%%%%%%%%%%%%%%%%%%%%%%%%%%%%
On the one hand, the results shown in Fig.~\ref{fig9}
supports our general conclusion that $\epsilon^p \sim R_{yy} K/y^2$ within a prefactor that
we cannot estimate from the theory. On the other hand the agreement is still not
perfect; the measured $\epsilon^p$ shows two maxima as a function of $y^+$. The first peak is related to the maximum of $R_{yy}K/y^2$ while the second peak
corresponds to the maximum of $\la R_{xx}^2 s_{xx}^2 \ra$. 
One of the central statements of the asymptotic theory \cite{additive} 
is that this term drops, by exact cancellation with another term, when
the MDR is approached in $\De \to \infty$.
This is not occurring yet in our simulations with low \Re~ and \De.  

%%%%% FIGURE 6 %%%%%%%%%%%%%%%%%%
\begin{figure}
\centering
\epsfig{width=.40\textwidth,file=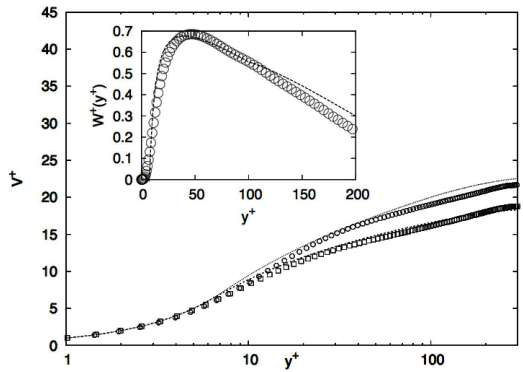}
\caption{ Comparison of the profiles  $V^+(y^+)$ and
$W^+(y^+)$ as computed from the balance equations (\ref{mom}) and(\ref{ene}) with 
the results of numerical simulations. The agreement is excellent.}
\label{fig6}
\end{figure}
%%%%%%%%%%%%%%%%%%%%%%%%%%%%%%%%%%

From our numerical simulations and the theoretical analysis we can also
state that $R_{yy}$ is linearly growing up to $y^+ \sim 80$. To see this,
estimate the effect of linear viscosity profile using the following equations \cite{05BDLP}:
\bea
\label{mom}
[1+\alpha(y-\delta)]S^+ + W^+ = 1\ , \\
\label{ene}
[1+\alpha F (y-\delta)]\frac{\Delta^2(\alpha)}{y^2} + \frac{1}{\kappa_k y} = S^+ \ .
\eea
In asymptotic conditions $F=1$, but here the factor $F$ takes into account that the effective slope of the linear viscosity
profile is somehow smaller for the energy balance equation than for the momentum equation.
According to our previous discussion on Fig.~(\ref{fig9}), we can estimate $F\approx 2$.
In (\ref{ene}) the term $\Delta(\alpha)$ was determined theoretically \cite{05BDLP}:
\be
\Delta(\alpha) = \frac{\delta}{1-\alpha\delta}
\ee
In our case the value of $\alpha$ is given by the relation
\be
\alpha = \frac{\nu_p}{\nu} \frac{dR_{yy}}{dy}
\ee
where the slope $dR_{yy}/dy$ is estimated from the numerical simulation.
Solving these simple coupled equations we present
in Fig.~(\ref{fig6}) the quantities $V^+(y^+)$ and
$W^+(y^+)$ respectively for the Newtonian flow and for the polymer laden flow. Both
figures agree well with the DNS.
%%%%%%%%%%%%%%%%%%%%%%%%%
\section{Conclusions}

The general philosophy behind our approach to drag reduction by additives 
is to consider the balance equations for mechanical momentum and turbulent energy,
and to analyze the predictions of these equations for the profiles of the 
relevant quantities, in particular the mean velocity at distance $y$
from the wall \cite{RMP}. The theory can be simplified in the asymptotic regime 
when \Re~ and \De~ are very large; there one finds universal profiles, in particular 
for the the mean velocity profile which becomes the universal MDR \cite{05BDLP}. 
For comparison with numerical simulations, where the drag reduction effect is 
rather limited due to small finite \Re and \De, 
one needs to analyze the balance equations with greater care, taking into 
account the non-asymptotic effects. This is what we have done in the present work, 
and our results are shown to agree well with the data obtained from direct numerical
simulations. 

%Thus the lesson is that one can analyze the asymptotic situation 
%with asymptotic theories, or low \Re~ situations with non-asymptotic theories. 
%The one thing that one cannot do is to analyze numerical simulations with 
%asymptotic theories.

\acknowledgments

ESCC acknowledges support by the Hong Kong Research Grants Council (CA05/06.SC01).
IP acknowledges partial support by the US-Israel Binational Science Foundation.


\begin{thebibliography}{99}

\bibitem{Virk}
P.S. Virk, 1975, AIChE J. {\bf 21}, 625; 
P. S. Virk, D. L. Wagger and E. Koury, 1996, ASME FED-
{\bf 237}, 261; 
P.S. Virk, D.C. Sherma and D.L. Wagger, 1997, AIChE J., {\bf 43}, 3257.

\bibitem{Bonn}
C. Wagner, Y. Amarouch\`ene, P. Doyle and D. Bonn, 2003,  Europhys. Lett. {\bf 64}, 823.

\bibitem{RMP}
  I. Procaccia, V. S. L'vov and R. Benzi,  ``Colloquium: Theory of Drag Reduction by Polymers in Wall Bounded Turbulence", Rev. of Mod. Phys., submitted Feb.15, 2007 Also:nlin.CD/0702034.
  
 \bibitem{fibres} 
 J.S. Paschkewitz, Y. Dubief, C.D. Dimitropoulos, E.S.G. Shaqfeh, and
P. Moin, J. Fluid Mech. {\bf 518}, 281 (2004).



\bibitem{DoiEdwards} 
M. Doi and S.F. Edwards, {\it The Theory of Polymer
Dynamics} (Oxford, 1988).


\bibitem{additive} R. Benzi, E.S.C. Ching, T.S. Lo, V.S. L'vov, and I. Procaccia,
Phys. Rev. E {\bf 72}, 016305 (2005).

\bibitem{04LPPT}
V.S. L'vov, A. Pomyalov, I. Procaccia and V. Tiberkevich, Phys. Rev. Lett.,  {\bf 92}  244503, (2004).

\bibitem{04DCLPPT}
cf. for example Fig. 4 in E. De Angelis, C.M. Casciola, V.S. L'vov, A. Pomyalov, I. Procaccia and
V. Tiberkevich, Phys. Rev. E, {\bf 70}, 055301 (2004).

\bibitem{05BDLP}
R. Benzi, E. De Angelis, V.S. L'vov and I. Procaccia, Phys. Rev. Lett., {\bf 95}, 194502 (2005).

\end{thebibliography}
\end{document}